# Electronic band structure, Fermi surface, structural and elastic properties of two polymorphs of MgFeSeO as possible new superconducting systems

I.R. Shein,* A.L. Ivanovskii

*Institute of Solid State Chemistry, Ural Branch of the Russian Academy of Sciences, 620990 Ekaterinburg, Russia*

**Abstract**

By means of DFT-based first-principles calculations, we examine two polymorphs of the newly synthesized 1111-like MgFeSeO as possible new superconducting systems. We have found that the polymorph with blocks [MgO], where Mg atoms are placed in the centers of $O_4$ tetrahedra, is dynamically unstable - unlike the ZrCuSiAs-type polymorph with oxygen atoms placed in the centers of $Mg_4$ tetrahedra. The characterization of this material covers the structural, elastic properties, electronic band structure, density of electronic states, and Fermi surface. Our calculations suggest that a high critical temperature for MgFeSeO may be achieved as a result of electron or hole doping through ion substitutions or through creation of lattice vacancies.

* *E-mail address:* shein@ihim.uran.ru

Since the discovery of superconductivity in the layered LaFeAsO$_{1-x}$F$_x$ with a critical temperature $T_C \sim 26K$ [1], intensive efforts have been focused on the search of new Fe-based superconductors (SCs), and today a lot of related iron-pnictide (Fe-*Pn*) SCs have been successfully prepared and investigated. These materials are united into several groups, namely so-called 111 (s.g. *P4/nmm*, such as LiFeAs), 122 (s.g. *I4/mmm*, such as *A*Fe$_2$*Pn*$_2$, *A* = alkaline earth metals), and 1111 (s.g. *P4/nmm*, such as *R*Fe*Pn*O, *R* = rare earths) systems, reviews [2-10]. Additionally, more complex multi-component 32225 (such as Sr$_3$Sc$_2$Fe$_2$As$_2$O$_5$) or 21113 phases (such as Sr$_2$*M*Fe*Pn*O$_3$, where *M* are Sc, Ti, V *etc.*, and their homologs such as Ca$_{n+2}$(Al,Ti)$_n$Fe$_2$As$_2$O$_z$; n = 2, 3 or 4) have been reported as parent phases for the new Fe-*Pn* SCs, reviews [2-10].

The family of related iron-chalcogenide (Fe-*Ch*) SCs, which includes 122 - like phases (s.g. *I4/mmm*, such as *A*$_x$Fe$_{2-y}$Se$_2$, *A* = alkaline metals) along with archetypical β-FeSe (11 phase, s.g. *P4/nmm*), is much more limited, reviews [11-14].

Very recently, the new tetragonal phase MgFeSeO [15] (for which at first the superconductivity with $T_C \sim 42K$ was reported) and which is isostructural with the aforementioned 1111 - like Fe-*Pn* SCs. As far as we know, MgFeSeO is the first Fe-containing 1111-like chalcogenide oxide: unlike a rich family of 1111-like pnictide oxides [9,10,16], only a limited number of related chalcogenide oxides (*M'MCh*O, where *M* = Bi or rare-earth metals, *M'* = Cu, Ag and *Ch* = S, Se, Te ) was known up to now, review [17]. Many of them are transparent *p*-type semiconductors, which have drawn attention as promising materials for various applications in optoelectronics [18-22].

In this Letter, using the DFT-based first-principles calculations, the electronic band structure, Fermi surface, structural and elastic properties of MgFeSeO, as a possible new superconducting system, have been examined.

Herein, we have considered two possible polymorphs of MgFeSeO, namely the system with the experimentally established "classical" ZrCuSiAs-type structure (like *R*Fe*Pn*O), when oxygen atoms are placed in the centers of Mg$_4$ tetrahedra [15] (further as MFSO-**I**), and the proposed [23] alternative polymorphic modification with "inverse" structure of blocks [Mg$_2$O$_2$], when Mg atoms are placed in the centers of O$_4$ tetrahedra (further as MFSO-**II)**, see Fig. 1.

In our studies, two complementary first-principles approaches are used. The dynamical stability of the polymorphs MFSO-**I,II** is examined by means of the Vienna *ab initio* simulation package (VASP) in the projector augmented waves (PAW) formalism [24,25]. Exchange and correlation are described in the form of generalized gradient correction (GGA) to exchange-correlation potential of Perdew, Burke, and Ernzerhof [26]. The geometry optimization is performed with the force cutoff of 2 meV/Å. Within this method, the structural and elastic properties for MgFeSeO are also evaluated. A further insight into the electronic properties of MgFeSeO is obtained by means of the full-potential method with mixed basis APW+lo (FLAPW) implemented in the WIEN2k suite of programs [27]. Exchange and correlation are also described in the form of GGA [26].

At the first stage, comparing the total energies $E_{tot}$ of MFSO-**I** *versus* MFSO-**II,** we found that $E_{tot}^{MFSO-II}$ is 0.08 eV/form.unit lower than that for MFSO-**I**. This is in accordance with the estimations [23]; therefore the authors [23] assume that the structure of MFSO-**II** is more energetically favorable.

However we continued the comparative **stability probing** of these polymorphs, using the condition of dynamical stability, when the Gibbs free energy of the crystal is in a local minimum with respect to small structural deformations. In terms of elastic constants ($C_{ij}$), this stability criterion for tetragonal crystals requires [28] that: $C_{11} > 0$, $C_{33} > 0$, $C_{44} > 0$, $C_{66} > 0$, $(C_{11} - C_{12}) > 0$, $(C_{11} + C_{33} - 2C_{13}) > 0$, and $\{2(C_{11} + C_{12}) + C_{33} + 4C_{13}\} > 0$. We have evaluated six independent elastic constants ($C_{11}$, $C_{12}$, $C_{13}$, $C_{33}$, $C_{44}$, and $C_{66}$) for polymorphs MFSO-**I,II** by calculating the stress tensors on different deformations applied to the equilibrium lattice of the tetragonal unit cell, whereupon the dependence between the resulting energy change and the deformation was determined. We found that for MFSO-**II** $C_{44} < 0$; thus, we can assert that this structure is unstable. On the contrary, for the experimentally reported [15] structure MFSO-**I** all of the constants $C_{ij}$ are positive and satisfy all of the aforementioned conditions; thus, this crystal is dynamically stable, and further we will focus on this system.

Next, let us discuss the **elastic properties** for MFSO-**I**. The calculated elastic constants $C_{ij}$ allow us to estimate numerically bulk ($B$), shear ($G$) and Young's ($Y$) moduli, as well as compressibility ($β$), Pugh's indicator ($G/B$ ratio), and Poisson's ratio ($v$), see for example [29]. The calculated values are: $B$ = 41.4 GPa, $β$ = 0.02416 GPa$^{-1}$, $G$ = 25.6 GPa, $Y$ = 63.7 GPa, $v$ = 0.244, and $G/B$ = 0.618. This suggests that the parameter limiting the mechanical stability of this material is the shear modulus ($B > G$). Besides, according to the Pugh's criterion [30], MFSO-**I** should behave in a brittle manner: $G/B > 0.57$. Additionally, our estimation of elastic anisotropy using the so-called universal index [31] $A^U = 5G_V/G_R + B_V/B_R - 6$ (where $G_{V,R}$ and $B_{V,R}$ are respectively the shear and bulk moduli within the Voigt (V) and Reuss (R) limits [29]; for isotropic crystals $A^U = 0$; deviations of $A^U$ from zero define the extent of crystal anisotropy) reveals that MFSO-**I** should demonstrate high elastic anisotropy ($A^U$ ~34.8). The obvious reason is a very weak inter-block bonding (of van-der-Waals type) in MFSO-**I,** see also below**.** Note that this reason explains also the essential softening of the examined MFSO-**I,** whose basic elastic moduli ($B$, $G$, and $Y$) decrease considerably (by more than 50%) in comparison with isostructural LaFeAsO, which exhibits ionic inter-block bonding [32].

As to the **electronic properties** of MFSO-**I,** within the FLAPW method we have calculated (for the optimized structural parameters: $a = b$ = 3.8120 Å, $c$ = 9.4014 Å, which are in reasonable agreement with the experiment [15]: $a = b$ = 3.7872 Å, $c$ = 9.2699 Å), the electronic band structure, densities of states (DOSs), and the Fermi surface (FS).

From Figs. 2 and 3, where the electronic bands and the Fermi surface of MFSO-**I** are depicted, we can see that the bands crossing the Fermi level yield a multi-sheet FS, which consists of two groups of quasi-two-dimensional sheets parallel to the $k_z$ direction, which include electron-like cylinders in the corners of the Brillouin zone and the hole-like cylinders along the Γ-Z line. A similar topology of FS was obtained for isostructural LaFeAsO [2-6]. From the densities of states (Fig. 2) we also see that the states near the Fermi level are formed exclusively by the Fe3d- and Se4p-orbitals of blocks [FeSe], whereas the blocks [MgO] are insulating.

What may be *the origin of superconductivity* of MgFeSeO with a high $T_C$ (~ 42K [15])? Since the experimental data [15] are very limited, for preliminary conclusions it is possible to use the empirical correlation [33] between $T_C$ of Fe-based SCs and the so-called anion height $z_a$ (with respect to Fe sheet), when the highest superconducting temperatures ($T_C$ > 40K) are achieved for $z_a$ ~ 1.38 Å. According to the experiment, for MgFeSeO $z_a$ ~ 1.52 Å [15]; our calculations give $z_a$ ~ 1.27 Å. The both values are far from the aforementioned "optimal" $z_a$. Another point, which casts doubts on the high $T_C$ for ideal stoichiometric MgFeSeO, comes from the estimation of $T_C$ within the simplified model [12]. Herein, the BCS-like expression for $T_C = 1.14\omega_D e^{-2/gN(E_F)}$ was employed, and, as the authors say [12], these estimations do not necessarily imply electron-phonon pairing, as $\omega_D$ may just denote the average frequency of any other possible Boson responsible for pairing interaction (e.g. spin fluctuations). Using the parameters $\omega_D$ and $g$ from Ref. [12] and the calculated value of the total DOS at the Fermi level $N(E_F)$ = 1.039 states/eV·cell, we obtain $T_C$ < 1K, which is far from the reported $T_C$ ~ 42K [15].

This allows us to assume that the synthesized sample [15] does not adopt the ideal stoichiometry MgFeSeO, but contains various crystal imperfections such as impurities or lattice vacancies, which act as electron- or hole dopants. Note that the majority of the parent phases of the Fe-based SCs show a high chemical flexibility to a large variety of constituent elements or to introduction of lattice vacancies. So, already for the first discovered Fe-based superconductor LaFeAsO$_{1-y}$F$_y$ it was revealed that the superconducting transition emerges after suppression of antiferromagnetic ordering in the parent phase LaFeAsO as a result of fluorine doping [1]. For 122 - like Fe-*Ch* SCs, the presence of Fe vacancies and their ordering also play an important role in regulating the properties of these systems [11,12,34]. Therefore the effects of hole- or electron doping on the properties of the new SC MgFeSeO are of interest.

From the estimations within the aforementioned model, the value of $N(E_F)$ should be about 4.36 states/eV·cell to achieve the reported $T_C$ ~ 42K [15]. For the stoichiometric MgFeSeO, the Fermi level is placed in the local minimum between two sharp DOS peaks (Fig. 2); thus, electron or hole doping can be a successful strategy for increasing $N(E_F)$. To simulate this effect, we used the obtained band picture for the "ideal" MFSO-**I** and shifted the Fermi level at -0.36 eV and +0.78 eV so that the required value of $N(E_F)$ is achieved in case of hole or electron doping, respectively. In Fig. 3, the corresponding FSs are depicted, which retain the characteristic quasi-two-dimensional topology typical of superconducting Fe-*Pn* and Fe-*Ch* materials. This gives additional hints that the experimentally observed high $T_C$ can be related to the presence of some imperfections in the synthesized sample of MgFeSeO.

Finally, let us discuss *the inter-atomic bonding* for the examined MFSO-**I**. One of the characteristic peculiarities of 1111-like pnictide oxides is the alternation of negatively charged conducting blocks [FeAs] with positively charged oxide blocks (so-called "charge reservoirs") providing the ionic bonding between the adjacent blocks [2-10]. On the contrary, for MgFeSeO, using the usual oxidation numbers of atoms: $Mg^{2+}$, $O^{2-}$, $Fe^{2+}$, and $Se^{2+}$, we obtain the ionic formula of this phase: $[Mg^{2+}O^{2-}]^0[Fe^{2+}Se^{2-}]^0$; this means the absence of inter-block charge transfer.

The common bonding picture for MgFeSeO illustrated by the electron density map in Fig. 1 shows that very anisotropic interactions arise in this layered crystal. Indeed, in blocks [MgO] and [FeSe], directional Mg-O and Fe-Se bonds take place owing to hybridization of the corresponding valence orbitals, see Fig. 2. Besides, the ionic interactions Mg-O and Fe-Se arise due to Mg → O and Fe →Se charge transfer. To estimate numerically the amount of electrons redistributed between various ions, we carried out a Bader's [35] analysis; the effective atomic charges $Q_{at}$ are $Mg^{0.303+}$, $Fe^{0.477+}$, $Se^{0.501-}$ and $O^{0.279-}$. Thus, the values of $Q_{at}$ become significantly smaller than within the ideal ionic model – owing to the aforementioned covalent effects. These results also show that the charge transfer between the adjacent blocks [MgO]/[FeSe] is very small ($\Delta Q \sim 0.02\ e$) confirming the assumption about very weak inter-block bonding (of the van-der-Waals type).

In summary, by means of two complementary DFT-based first-principles approaches, two polymorphs of the 1111-like MgFeSeO, as a possible new superconducting system, were examined. We found that the polymorph MFSO-**II** with blocks [MgO], where Mg atoms are placed in the centers of $O_4$ tetrahedra, is dynamically unstable - unlike the ZrCuSiAs-type polymorph MFSO-**I,** with oxygen atoms placed in the centers of $Mg_4$ tetrahedra. For this material, the structural, elastic, and electronic properties have been predicted.

We found that MFSO-**I** is a soft material with rather small bulk, shear and Young's moduli, which behaves in a brittle manner and exhibits essential elastic anisotropy. These peculiarities are related to very anisotropic inter-atomic bonding in this 2D-like crystal, where strong covalent-ionic bonds take place inside blocks [MgO] and [FeSe], while between the adjacent blocks [MgO]/[FeSe] very weak (van-der-Waals type) bonds emerge.

The near-Fermi states of MFSO-**I** are formed exclusively by the Fe3d- and Se4p-orbitals of blocks [FeSe], whereas the blocks [MgO] are insulating. The Fermi surface is of a multi-sheet type and consists of two groups of hole- and electron-like cylinders parallel to the $k_z$ direction.

At the same time, our calculations suggest that according to available indicators the high critical temperature for the ideal stoichiometric MgFeSeO [15] seems doubtful and may be due to electron or hole doping through ion substitutions or through creation of lattice vacancies. Certainly, this issue requires further experimental and theoretical efforts.

_________________

1.  Y. Kamihara, T. Watanabe, M. Hirano, and H. Hosono, J. Amer. Chem. Soc. **130**, 3296 (2008).
2.  M. V. Sadovskii, Physics - Uspekhi **51**, 1201 (2008).
3.  L. Ivanovskii, Physics - Uspekhi **51**, 1229 (2008).
4.  Z. A. Ren, and Z. X. Zhao, Adv. Mater. **21**, 4584 (2009).
5.  J. Paglione, and R. L. Greene, Nature Phys. **6**, 645 (2010).
6.  D. C. Johnson, Adv. Phys. **59**, 803 (2010).

7. L. Ivanovskii, Russ. Chem. Rev. **79**, 1 (2010).
8. G. R. Stewart, Rev. Mod. Phys. **83**, 1589 (2011).
9. D. Johrendt, J. Mater. Chem. **21**, 13726 (2011).
10. D. Johrendt, H. Hosono, R.D. Hoffmann, and R. Pöttgen, Zisch. Kristallogr. **226**, 435 (2011).
11. L. Ivanovskii, Physica C **471**, 409 (2011).
12. M. V. Sadovskii, E. Z. Kuchinskii, and I. A. Nekrasov, J. Magn. Magn. Mater. **324**, 3481 (2012).
13. D. J. Singh, Sci. Technol. Adv. Mater. **13**, 054304 (2012).
14. Fujitsu, S. Matsuishi, and H. Hosono, Intern. Mater. Rev. **57**, 311 (2012).
15. X. F. Lu, N. Z.Wang, G. H. Zhang, et al., arXiv:1309.3833.
16. C. Ozawa, and S. M. Kauzlarich, Sci. Technol. Adv. Mater. **9**, 033003 (2008).
17. R. Pöttgen, and D. Johrendt, Ztsch. Naturforsch. B: **63**, 1135 (2008).
18. K. Ueda, H. Hiramatsu, M. Hirano, et al., Thin Solid Films **496**, 8 (2008).
19. H. Hosono, Thin Solid Films **515**, 6000 (2007).
20. H. Hiramatsu, H. Yanagi, T. Kamiya, et al., Chem. Mater. **20**, 326 (2008).
21. H. Hosono, Physica C **469**, 314 (2009).
22. V. V. Bannikov, I. R. Shein, and A.L. Ivanovskii, Solid State Sci. **14**, 89 (2012).
23. K. Liu, M. Gao, Z. Y. Lu, and T. Xiang, arXiv:1309.5079.
24. G. Kresse, and D. Joubert, Phys. Rev. B **59**, 1758 (1999).
25. G. Kresse, and J. Furthmuller, Phys. Rev. B **54**, 11169 (1996).
26. J. P. Perdew, S. Burke, and M. Ernzerhof, Phys. Rev. Lett. **77**, 3865 (1996).
27. P. Blaha, K. Schwarz, G. Madsen, et al., WIEN2k, *An Augmented Plane Wave Plus Local Orbitals Program for Calculating Crystal Properties*, Vienna University of Technology, Vienna, 2001.
28. M. Born, and K. Huang, *Dynamical Theory of Crystal Lattices*, Clarendon, Oxford, 1956.
29. L. Ivanovskii, Progr. Mater. Sci. **57**, 184 (2012).
30. F. Pugh, Phil. Mag. **45**, 823 (1954).
31. S. I. Ranganathan, and M. Ostoja-Starzewski, Phys. Rev. Lett. **101**, 055504 (2008).
32. R. Shein and A.L. Ivanovskii, Scripta Materialia **59**, 1099 (2008).
33. Y. Mizuguhci, Y. Hara, K. Deguchi, et al., Supercond. Sci. Technol. **23**, 054013 (2010).
34. D. X. Mou, L. Zhao, and X. J. Zhou, Frontiers Phys. **6**, 410 (2011).
35. R. Bader, A*toms in Molecules: A Quantum Theory*, University Press, New York, 1990.

**FIGURES**

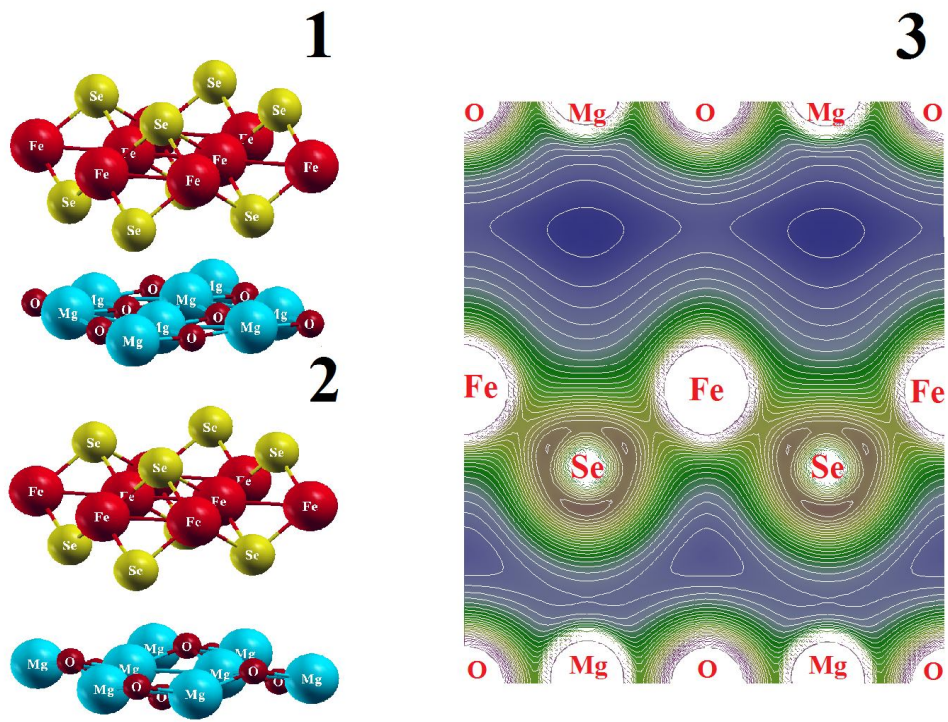

Fig. 1. Atomic structures of two possible polymorphs of MgFeSeO with various types of blocks [MgO]: MFSO-**I** (1) and MFSO-**II** (2), and the electron density map for MFSO-**I** (3).

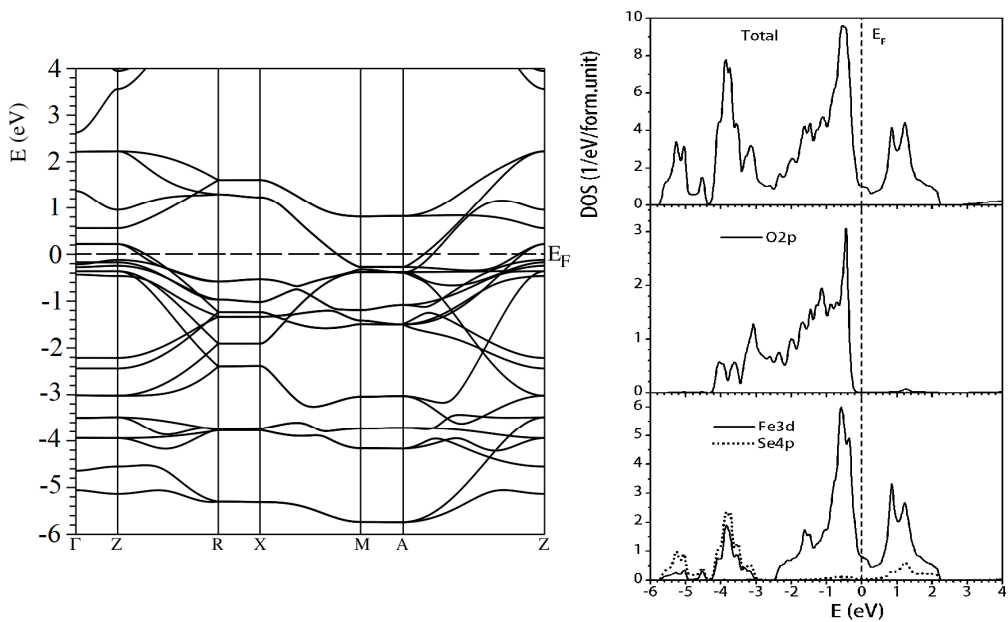

Fig. 2. Electronic bands (o*n the right*) and total and atomic-projected densities of states for MFSO-**I** (*on the left*).

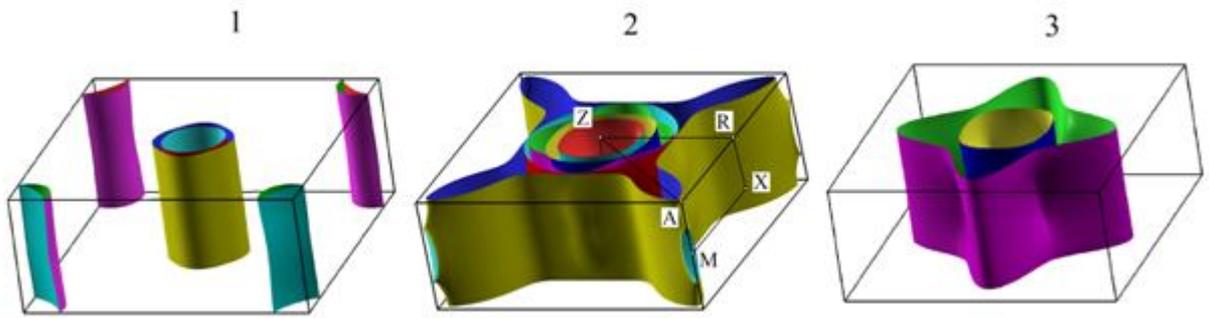

Fig. 3. Fermi surfaces for the ideal stoichiometric MFSO-I (1) and for this system doped with holes (2) and electrons (3), *see the text*.